%% file: main.tex
\documentclass[fleqn,usenatbib,margin=1in]{mnras}

\usepackage{newtxtext,newtxmath}

\usepackage[T1]{fontenc}

\DeclareRobustCommand{\VAN}[3]{#2}
\let\VANthebibliography\thebibliography
\def\thebibliography{\DeclareRobustCommand{\VAN}[3]{##3}\VANthebibliography}

\usepackage{graphics}
\usepackage{listings}
\usepackage{enumitem}
\usepackage{url}
\usepackage{appendix}
\usepackage{xcolor}

\usepackage{hyperref}
\usepackage{mathtools}

\usepackage[super]{nth}

\begin{document}

\author[Richard M. Feder]{Richard M. Feder$^{1,2}$\thanks{E-mail: rmfeder@berkeley.edu},{Martin White$^{1,2}$} \\
$^{1}$Berkeley Center for Cosmological Physics, University of California, Berkeley, CA 94720, USA\\
$^{2}$Lawrence Berkeley National Laboratory, Berkeley, California 94720, USA
}

\title{Angular BAO Forecasts for the IBIS Medium-Band Survey}
\maketitle

\begin{abstract}
Ongoing and near-future spectroscopic surveys, such as DESI, DESI-II and Spec-S5, rely on imaging-based selections to construct uniform, three-dimensional tracers of large-scale structure. While spectroscopic data from these surveys constrains the baryonic acoustic oscillation (BAO) feature with high precision, the imaging surveys used for target selection can provide useful information on the angular diameter distance $D_A(z)$. In this work we explore the feasibility of angular BAO measurements for the Intermediate-Band Imaging Survey (IBIS) using recent constraints on clustering from a pilot survey spanning $2.2<z<3.5$. Through Fisher forecasts, we find that a 5000 deg$^2$ survey of LAEs with realistic bias, a tracer density of $2\times 10^{-4}$ (h/Mpc)$^3$ and interloper fraction $f_{\rm int}=10\%$ can constrain the BAO dilation parameter $\alpha$ at $z_{\rm eff}=2.8$ with a precision of 2.6\%, with dependence on the sample properties that is consistent with shot noise-dominated measurements. We then explore medium-band survey specifications for the planned Stage-V Spectroscopic Instrument (Spec-S5) and beyond, demonstrating the potential for precise high-redshift BAO measurements. Our forecasts motivate early measurements of BAO from these imaging surveys, which may inform later spectroscopic analyses.
\end{abstract}

\input{introduction}
\input{angular_bao}

\input{survey_spec}

\input{forecasts}
\input{conclusion}

\bibliographystyle{aasjournal}
\bibliography{references}

\section*{Software}
This work made use of the packages \texttt{matplotlib}, \texttt{numpy}, and \texttt{velocileptors}.

\end{document}

%% file: introduction.tex
\section{Introduction}

Baryon acoustic oscillations (BAO) in the large-scale structure (LSS) of galaxies provide an excellent standard ruler for mapping the expansion history across cosmic time \citep{eisenstein98}. Measurements of the BAO feature in galaxy clustering constrain the angular diameter distance $D_A(z)$ and Hubble rate $H(z)$, enabling precision tests of the $\Lambda$CDM cosmological model \citep{lcdm_review}. Since the first BAO detections in 2005 using the Sloan Digital Sky Survey \citep[SDSS;][]{eisenstein_bao} and the 2-degree Field Galaxy Redshift Survey \citep[2dFGRS;][]{cole_bao}, both spectroscopic and photometric redshift surveys have utilized BAO as a cornerstone probe of cosmic acceleration \citep{riess98, perlmutter99}, anchoring and complementing constraints from the cosmic microwave background and Type Ia supernovae.

The Dark Energy Spectroscopic Instrument (DESI) has recently delivered BAO measurements spanning $z \simeq 0.1$--$2.5$ using luminous red galaxies, emission-line galaxies, quasars, and through absorption with the Lyman-$\alpha$ forest \citep{desi_dr2_bao}. When combined with supernova data, these results have exhibited growing tensions with a flat $\Lambda$CDM expansion history. Extending BAO observations to higher redshift offers access to the matter-dominated epoch, in which $D_A(z)$ and $H(z)$ are primarily sensitive to $\Omega_m$ and $\Omega_k$. This long lever arm in redshift improves constraints on curvature and dark energy, while enabling stringent internal consistency tests of the standard cosmological model \citep{ferraro_snowmass, sailer_highz}. 

High-redshift BAO surveys face the observational challenge of assembling sufficiently large, well-characterized tracer samples, but benefit from cleaner theoretical modeling: weaker structure growth at early times reduces the non-linear damping of the BAO signal, with smaller effective $\Sigma_{\perp}$ and $\Sigma_{\parallel}$ than at lower redshifts. Suitable tracer populations include Lyman-$\alpha$ emitters \citep[LAEs;][]{lyman_alpha} and Lyman break galaxies \citep[LBGs;][]{lbgs}, whose strong rest-frame ultraviolet features enable efficient selection at $z \gtrsim 2$.

DESI-II \citep{schlegel_desi2}, planned to begin operations in the late-2020s, will target LAEs and LBGs across $2.3 < z < 3.5$. While the exact survey specifications will be finalized in future work, the Intermediate-Band Imaging Survey (IBIS) will be a central part of the imaging used for target selection of these tracers, in combination with broad-band imaging from \emph{Subaru} and \emph{Rubin} LSST. IBIS, which deploys a set of medium-band filters on the Dark Energy Camera \citep[DECam;][]{decam} (DECam) at the Blanco telescope in Chile, can be used to isolate spectral breaks and emission lines.

Pathfinder work by \citet{ebina_ibis} combined IBIS data with DESI spectroscopy over a targeted $\sim$10 deg$^2$ region to study the clustering of LAEs and LBGs\footnote{The clustering of LBGs from broad-band selection was studied by \citet{RK24}.}, joining other narrowband imaging efforts such as the One-hundred-square-degree DECam Imaging in Narrowbands \citep[ODIN;][]{Lee24,white_odin_24,Herrera25}, the VIMOS VLT Deep Survey \citep[VVDS;][]{vvds}, and Hobby-Eberly Telescope Dark Energy Experiment \citep[HETDEX;][]{hetdex_gebhardt, hetdex}. Furthermore, \cite{ebina_ibis} demonstrated that photometric redshifts derived from medium-band photometry achieve redshift precision of $\sigma_z/(1+z) \sim 0.05$, as validated with DESI spectroscopic follow-up. 

While DESI-II spectroscopy is expected to yield sub-percent BAO measurements at these redshifts, it is natural to ask whether the DESI-II imaging survey alone could provide meaningful BAO constraints prior to spectroscopic completion. Prior work using Sloan Digital Sky Survey (SDSS) imaging has demonstrated that photometric samples contain useful cosmological and baryonic oscillation information, both in 3D and angular power spectra \citep{blake2007, padmanaban_2007, seo_ho_white_2012}. Angular BAO measurements have since improved with  imaging data from the Dark Energy Survey \citep[DES;][]{des_y3_bao, des_y6_bao}, with the most recent DES-Y6 measurements reaching a precision of 2.1\% at $z_{\rm eff}=0.83$. 
By using sufficiently fine redshift binning, one can mitigate BAO smearing from projection and can obtain clean transverse BAO measurements, recently highlighted by analysis with SDSS quasars \citep{SDSS_quasar_transverse_BAO} using redshifts $1.72<z<1.73$. 

The benefits of such approaches to transverse BAO measurements rely on relatively precise redshifts and accurate $dN/dz$ determinations, which are best achieved through narrow-band imaging and/or spectroscopy. Although measurements of the anisotropic BAO are precluded by the relatively large photometric redshift errors from IBIS+broad-band photometry that source strong radial damping, the achieved photo-$z$ precision is sufficient to pursue angular BAO measurements that benefit from narrow redshift bins. Beyond providing a preliminary measurement of the BAO signal at $z>2$, angular clustering with IBIS can act as a forward scout for the spectroscopic analysis: calibrating photometric selection functions, identifying the redshift bins and tracer populations that deliver the highest cosmological leverage, and diagnosing any large-scale systematics that may impact subsequent 3D measurements. Early results from the imaging survey can therefore guide target prioritization and survey strategy for DESI-II spectroscopy, maximizing its efficiency in achieving sub-percent BAO precision.

In this work, we forecast the angular BAO sensitivity from medium-band-selected LAEs obtained with an IBIS-like sample, using angular auto- and cross-power spectra across five redshift bins spanning $2.2<z<3.5$ to recover baryonic oscillation information. We vary survey area, tracer number density, and interloper fraction to explore their impacts on BAO precision.
We begin in \S \ref{sec:bao_modl} by outlining the angular BAO modeling used for our forecasts. After describing our fiducial IBIS survey specifications in \S \ref{sec:survey_spec}, we present our forecasts in \S \ref{sec:forecasts} and their dependence on survey configuration. In \S \ref{sec:stageV} we look ahead to Stage-V surveys and associated medium-band imaging campaigns, which will enable precise measurements of the BAO scale up to $z=3.5$ and potentially up to $z=5$, absent full spectroscopic follow-up. We conclude in \S \ref{sec:conclusion}.


Throughout this work we assume the \emph{Planck} 2018 best-fit cosmological parameters \citep{planck18}.

%% file: angular_bao.tex
\section{BAO modeling}
\label{sec:bao_modl}

Our forecasts are derived by computing the angular galaxy power spectra,
$C_\ell^{ij}$, between redshift bins $i$ and $j$,
and propagating the sensitivity of these spectra to a dilation parameter $\alpha$
that rescales the BAO scale. We incorporate and marginalize over a set of smooth spline functions when reporting constraints on $\alpha$, which have been demonstrated in \cite{chen_bao_desi24} as an effective nuisance model parameterization for handling residual broadband contributions. This approach has been used by the DESI collaboration for recent BAO measurements \citep{desi_dr2_bao} in both configuration space and Fourier space.

\subsection{Redshift-space galaxy power spectrum}

We begin by expressing the linear matter power spectrum as
\begin{equation}
P_{\rm m}(k,z) \;=\; P_{\rm nw}(k,z) + P_{\rm w}(k,z)
\end{equation}
where $P_{\rm nw}$ denotes the broadband (``no-wiggle") spectrum and $P_{\rm w}(k,z)$ isolates the BAO oscillations (``wiggle" component). We use a wiggle/no-wiggle decomposition implemented from the code \texttt{velocileptors} \citep{velocileptors}, applying a Savitzky-Golay filter to the power spectrum with wiggles.


The galaxy power spectrum includes redshift-space distortions (RSD) and other non-linear damping effects. For two galaxy samples with linear biases $b_i$ and $b_j$, we express the redshift-space, no-wiggle power spectrum as
\begin{align}
P_{g,\text{nw}}^{ij}(k_\perp, k_\parallel, z) &= \sqrt{A_i(k, \mu) \cdot A_j(k, \mu)} \cdot P_{\text{nw}}(k, z), 
\end{align}
where 
\begin{equation}
    A_i(k, \mu)=\left[b_i D(z_i)(1+\beta \mu^2)\right]^2 D_{\text{FoG}}(k_{\parallel}).
\end{equation}
Here, $D(z_i)$ is the linear growth factor, $\beta_i = f(z_i)/b_i$ is the Kaiser RSD parameter, $k=\sqrt{k_{\perp}^2 + k_{\parallel}^2}$ and $\mu = k_\parallel/k$ is the cosine of the angle to the line of sight. We model the Finger-of-God (FoG) damping from small-scale random velocities as 
\begin{equation}
D_{\text{FoG}, i}(k_\parallel) = \exp\left[-k_\parallel^2 \sigma_{p,i}^2\right]
\end{equation}
where $\sigma_{p,i}$ is the pairwise velocity dispersion for the host halos in bin $i$\footnote{We confirm that replacing our Gaussian FoG damping model with a Lorentzian (i.e., $D_{\text{FoG}}(k_{\parallel}) = \left[ 1 + (\sigma_p k_{\parallel})^2\right]^{-1}$) has no appreciable impact on our results.}.


The wiggle component has additional non-linear damping due to transverse and line-of-sight structure. Namely,
\begin{align}
P_{g,\text{w}}^{ij}(k_\perp, k_\parallel, z) &= \sqrt{B_i(k,\mu)\cdot B_j(k,\mu)} \cdot P_{\text{w}}(k,z), 
\end{align}
where $B_i(k,\mu) = A_i(k, \mu) \times D_{\text{BAO},i}(k,\mu)$ and 
\begin{equation}
D_{\text{BAO}, i}(k, \mu) = \exp\left[-\frac{1}{2} k^2 \left( \Sigma_{\perp}^2(1-\mu^2) + \Sigma_{\parallel}^2\mu^2 \right)\right].
\end{equation}
The level of BAO suppression is characterized by parameters $\Sigma_{\perp}$ and $\Sigma_{\parallel}$, where $\Sigma_{\perp} = \Sigma_0 D(z)$, $\Sigma_{\parallel} = \Sigma_0 D(z)(1+f(z))$, and $\Sigma_0$ is the 1D linear theory displacement dispersion,
\begin{equation}
    \Sigma_0^2 = \int \frac{dk}{6\pi^2}P_{\rm lin}(k,z=0) \simeq [5.87 D(z) h^{-1} \rm Mpc]^2,
\end{equation}
where in the second equality we assume the growth factor is normalized such that $D(z=0)=1$. For the redshifts in consideration, $\Sigma_0$ ranges from $1.6-2.2$ h$^{-1}$ Mpc, which translates to angular scales $\ell_{nl} \gtrsim 2500$, meaning the effects of damping should be small for the range of scales considered in our forecasts. Likewise, the line-of-sight damping can be recast in terms of the effective redshift error, 
\begin{equation}
    \frac{\delta z}{1+z} = \frac{\Sigma_0(1+f)H(z)}{c(1+z)},
\end{equation}
which, for our redshift range of interest, corresponds to $\delta z/1+z < 0.08\%$. In other words, the radial damping is significantly smaller than the thickness of the shells we project over ($\Delta z/1+z \approx 5\%$).

We do not attempt to treat BAO reconstruction for the photometric samples considered here -- the large photometric redshift uncertainties remove radial modes and limit estimates of the displacement field needed for reconstruction. Moreover, the intrinsic BAO damping at $z\gtrsim 2.3$ is already small, such that potential gains from reconstruction are less pronounced than at lower $z$.



We also note that the effect of non-linear clustering can shift the BAO feature in redshift space, however for $z > 2$ and for tracers with $b \sim 2-3$ the expected shifts lead to $\Delta \alpha_{\parallel} \leq 0.3\%$ and $\Delta \alpha_{\perp} \leq 0.2\%$ \citep{chen_bao_desi24}. We will find that these shifts are negligible compared to the expected precision of the measurement. Likewise, we neglect the impact of magnification bias, which, for the explored survey configurations, should have little impact on our BAO forecasts.


\subsection{Angular power spectrum}

Given that the radial width of our redshift bins $\Delta\chi$ is only a few times larger than the BAO scale, the standard Limber approximation is insufficient as it considers only the $k_\parallel\approx 0$ contributions. We instead employ the plane-parallel (or flat-sky) approximation but integrate over the line-of-sight wavenumber $k_\parallel$: 
\begin{equation}
C_\ell^{g_i g_j} = \frac{1}{\pi \chi_i \chi_j} \int dk_\parallel \, P_{\text{g}}^{ij}(k_\perp, k_\parallel) \, |\tilde{\phi}_i(k_\parallel)| \, |\tilde{\phi}_j(k_\parallel)|,
\label{eq:cell_flatsky}
\end{equation}
where we define integration limits $k_{\parallel}^{\rm min}=\pi/\Delta\chi \approx 0.015$ h Mpc$^{-1}$ and $k_{\parallel}^{\rm max} = 1.0$ h Mpc$^{-1}$. In the above equation, $\chi_i$ is the comoving distance to the center of bin $i$, and the transverse wavenumber is related to the multipole by $k_\perp = \ell/\sqrt{\chi_i \chi_j}$. The term $\tilde{\phi}_i(k_\parallel)$ is the one-dimensional Fourier transform of the radial window function $\phi_i(\chi)$, which is derived from the normalized redshift distribution via $\phi_i(\chi) = (dN_i/dz)(dz/d\chi)$.

\begin{figure}
    \centering
    \includegraphics[width=\linewidth]{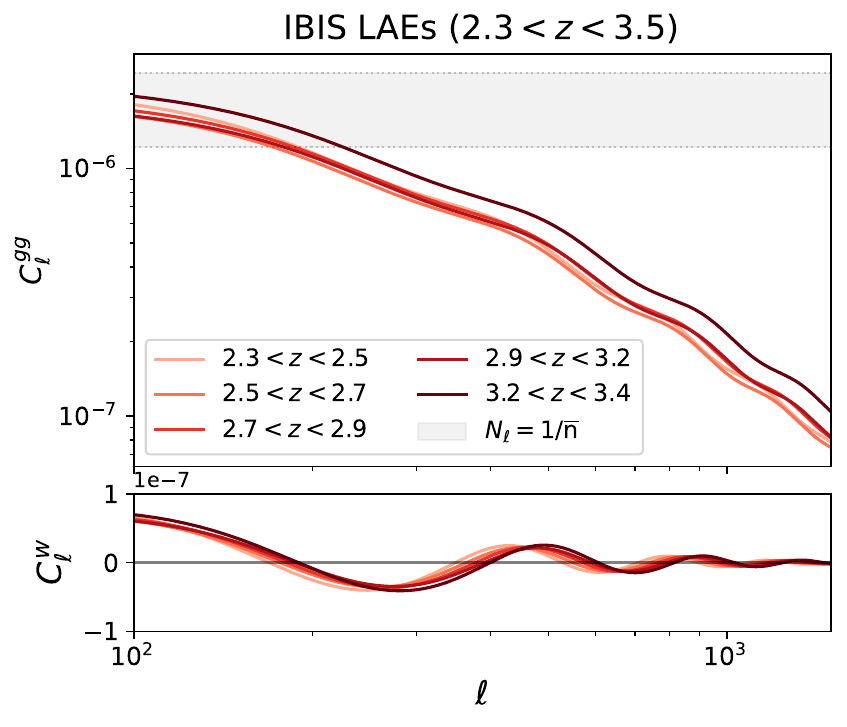}
    \caption{Top: LAE angular auto-power spectra, taking $b(z)=2.0-2.5$ as per Table \ref{tab:survey_specs}. The grey shaded region indicates the range of shot noise levels explored in our forecasts (see \S \ref{sec:survey_spec}). Bottom: residual BAO wiggle components. We restrict our analysis to $\ell > 100$, corresponding to $k\gtrsim 0.015$ at $z=3.0$.}
    \label{fig:c_ells}
\end{figure}

\subsection{BAO dilation parameter \texorpdfstring{$\alpha$}{alpha}}

The BAO scale acts as a standard ruler in the galaxy clustering signal,
with a characteristic comoving separation given by the sound horizon at
the drag epoch, $r_d$. A shift in the measured BAO feature relative to the fiducial model can be parameterized by the dimensionless dilation parameter $\alpha$, defined so that $\alpha = 1$ for the fiducial cosmology.

In Fourier space, a pure BAO dilation rescales wavenumbers $k$ in the oscillatory part of the matter power spectrum:
\begin{equation}
P_{\mathrm{m}}(k,z;\alpha) =
P_{\mathrm{nw}}(k,z)
+ P_{\mathrm{w}}\!\left(\frac{k}{\alpha}, z\right).
\label{eq:bao_alpha_def}
\end{equation}
We employ this phenomenological scaling, which is then projected into the angular power spectra $C_\ell^{ij}$ via Eq.~\eqref{eq:cell_flatsky}. A measurement of $\alpha$ deviating from a fiducial value of 1 corresponds to a measurement of the angular diameter distance $D_A(z)$ relative to the sound horizon,
\begin{equation}
\alpha = \frac{(D_A(z)/r_d)}{(D_A(z)/r_d)_{\rm fid}}.
\end{equation}

\subsection{Spline broadband removal}
\label{sec:splines}

We include nuisance parameters describing the cubic spline model introduced in \cite{chen_bao_desi24}. This basis allows for flexibility in the broadband modeling while also lacking capacity to reproduce the BAO signal itself. Our model takes the following form
\begin{equation}
\mathcal{D}_\ell^{ij} \;\approx\; \sum_{m} \beta_m^{ij} \; S_m(\ell),
\end{equation}
where $\beta_m^{ij}$ are linear nuisance coefficients, $S_m(\ell)$ are cubic spline functions, and the sum is over spline knots, which are spaced by $\Delta\ell_{\rm spline}$. 
The angular scale for BAO corresponds to 
$\ell_{BAO} \approx \chi(z)/r_d$, so by choosing a sufficiently large $\Delta \ell_{\textrm{spline}}$ we avoid removing true BAO in the broadband separation step.

For the redshifts in consideration ($2.2<z<3.5$), $\ell_{BAO} \approx 120-150$, though we note that BAO smearing from projection within redshift bins can spoil prescriptions for $\Delta \ell_{\textrm{spline}}$ based on $\ell_{BAO}$ and its harmonics alone. Through numerical tests in which we fit spline basis functions with varying $\Delta \ell_{\rm spline}$ to the BAO wiggle-only component spectra, we find that BAO signal loss is minimized for choices $\Delta \ell_{\rm spline}\geq 350$, and so we fix $\Delta \ell_{\rm spline}=350$ for our forecasts. We confirm that our degradation in forecast $\sigma(\alpha_{BAO})$ is also minimized for this choice of $\Delta \ell_{\rm spline}$ -- however, the precise tradeoffs between residual broadband structure and reduction of true BAO signal are beyond the scope of this work, and will be best understood through realistic mock recovery tests utilized in a real analysis.


\subsection{Fisher matrix}

We forecast constraints using the Fisher matrix formalism. Our model for each galaxy angular power spectrum is composed of BAO wiggles and a smooth broadband component described by a sum over spline basis functions,
\begin{equation}
C_{\ell, \mathrm{model}}^{g_i g_j}(\alpha, \boldsymbol{\beta}^{ij}) =
C_{\ell, \mathrm{w}}^{g_i g_j}(\alpha) +
\mathcal{D}_{\ell}^{ij}.
\label{eq:model}
\end{equation}

For parameters $p_a \in \{\alpha,\beta_m^{ij}\}$, we compute the Fisher matrix by summing information over multipoles $\ell$:
\begin{equation}
F_{ab} = \sum_{\ell=\ell_{\mathrm{min}}}^{\ell_{\mathrm{max}}}
\left(\frac{\partial \mathbf{C}_{\ell, \mathrm{model}}}{\partial p_a}\right)^T
\mathbf{Cov}_\ell^{-1}
\left(\frac{\partial \mathbf{C}_{\ell, \mathrm{model}}}{\partial p_b}\right),
\label{eq:fisher_vec}
\end{equation}
for which we assume a Gaussian covariance for the measured power spectra,
\begin{equation}
\mathrm{Cov}\left[C_\ell^{ij}, C_\ell^{m n}\right]
 = \frac{1}{(2\ell+1)f_{\rm sky}}
 \left[\tilde{C}_\ell^{im}\tilde{C}_\ell^{jn} +
\tilde{C}_\ell^{in}\tilde{C}_\ell^{jm}\right].
\end{equation}
The term $\tilde{C}_\ell^{ij}$ represents the total \textit{observed} power spectrum, which is the sum of the fiducial cosmological signal and the noise:
\begin{equation}
\tilde{C}_\ell^{ij} = C_{\ell}^{g_i g_j} + N_\ell^{ij}.
\end{equation}
We assume the noise spectra are dominated by galaxy shot noise, i.e., $N_\ell^{ij} = \delta_{ij} / \bar{n}_i^{\rm 2D}$. 

Because the spline parameters enter linearly, their derivatives are simply the basis functions:
\begin{equation}
\frac{\partial \mathcal{D}_\ell^{ij}}{\partial \beta_m^{ij}} = S_m(\ell),
\end{equation}
while the derivative with respect to $\alpha$ only involves the BAO wiggle component and is computed numerically using a central finite difference:
\begin{equation}
\frac{\partial C_{\ell,\mathrm{w}}^{ij}}{\partial \alpha}
\approx
\frac{
C_{\ell,\mathrm{w}}^{ij}(\alpha_0+\delta\alpha) -
C_{\ell,\mathrm{w}}^{ij}(\alpha_0-\delta\alpha)
}{2\,\delta\alpha}.
\label{eq:cell_deriv}
\end{equation}

Let the full Fisher matrix have the block form
\begin{equation}
\mathbf{F} =
\begin{pmatrix}
F_{aa} & F_{ab} \\
F_{ba} & F_{bb}
\end{pmatrix}
\end{equation}
where $a$ denotes our parameter of interest ($\alpha$) and $b$ the set of spline parameters ($\lbrace\beta_{m}^{ij}\rbrace$).
Marginalizing over $\lbrace\beta_{m}^{ij}\rbrace$ leads to our final constraints on $\alpha$:
\begin{equation}
F_{aa}^{\mathrm{marg}}
= F_{aa} - F_{ab}\,F_{bb}^{-1}\,F_{ba}.
\label{eq:schur}
\end{equation}
The forecasted $1\sigma$ uncertainty on $\alpha$ is then given by $\sigma_{\alpha} = (\sqrt{F_{aa}^{\rm{marg}}})^{-1}$.

%% file: survey_spec.tex
\section{IBIS survey specifications}
\label{sec:survey_spec}

We summarize our survey specifications in Table~\ref{tab:survey_specs}. Our fiducial survey configuration largely follows the IBIS medium-band galaxy sample clustering measurements presented in \cite{ebina_ibis}, spanning the redshift
range $2.3 < z < 3.4$ in five bins defined by filters of width $\Delta \lambda \approx 250$ \AA, with partial overlap between filters from matched cut-on/cut-off wavelengths. We model the galaxy $dN/dz$ in each bin following each filter's relative transmission, assuming a LAE medium-band excess selection strategy. While LBGs will be targeted with the same imaging data, their redshift distribution is typically broader, which leads to stronger smearing of the BAO feature after projection, and so we restrict our forecasts to LAEs. We assume a galaxy bias model that increases linearly across the range $2.3 < z < 3.5$ from $b=2.0$ to $b=2.5$. Given the relatively high bias of the assumed LAE population, we do not pursue any dedicated HOD modeling in our forecasts.



We assume a constant comoving galaxy density as a function of redshift, which we vary in four bins spanning $\overline{n}=[1.5-3.0]\times 10^{-4}$ $h^3$ Mpc$^{-3}$. These densities are slightly higher on average compared to the samples used in \cite{ebina_ibis}, reflecting anticipated improvements in selection efficiency for the full survey. We convert 3D galaxy densities to projected 2D densities, which range from 625 to 1250 deg$^{-2}$ for the total LAE sample. Lastly, we adopt a fiducial multipole range $100 < \ell < 1500$, chosen in order to recover the majority of the BAO signal across our redshift samples, which we elaborate on in \S \ref{sec:forecasts}.



\begin{table}
\centering
\caption{Fiducial survey specifications for our IBIS BAO forecasts.}
\label{tab:survey_specs}
\begin{tabular}{ccccc}
\hline
Redshift bin & $z_{\rm mean}$ & $b_g$ & $\bar{n}_{\rm LAE}$ & Surface density \\
& & & [$10^{-4} h^3\,{\rm Mpc}^{-3}$] & [deg$^{-2}$] \\
\hline
$[2.26, 2.56]$ & 2.41 & 2.0 & $1.5-3.0$ & $125-250$ \\
$[2.47, 2.77]$ & 2.62 & 2.1 & $1.5-3.0$ & $125-250$\\
$[2.68, 2.98]$ & 2.83 & 2.2 & $1.5-3.0$ & $125-250$\\
$[2.89, 3.20]$ & 3.05 & 2.3 & $1.5-3.0$ & $125-250$\\
$[3.10, 3.41]$ & 3.26 & 2.5 & $1.5-3.0$ & $125-250$\\
\hline
\end{tabular}
\end{table}

%% file: forecasts.tex
\section{Results}
\label{sec:forecasts}

\subsection{Fiducial IBIS forecasts}

\begin{figure}
    \centering
    \includegraphics[width=\linewidth]{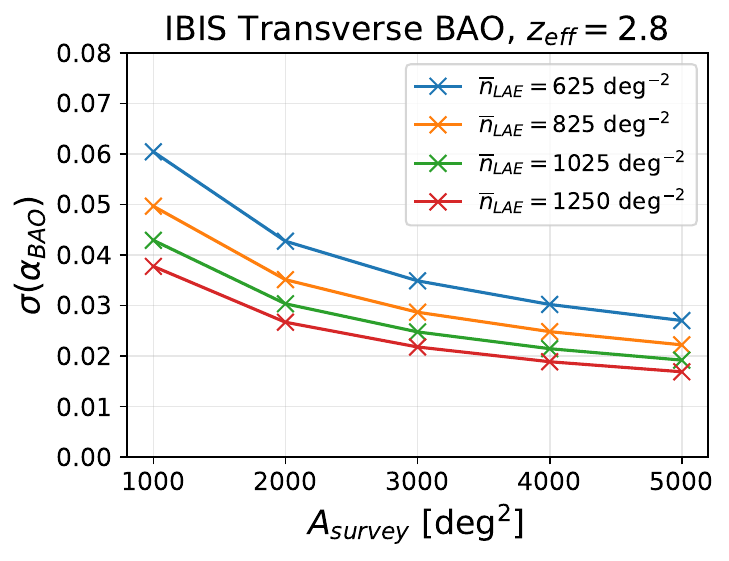}
    \caption{Forecast uncertainty on $\alpha_{\rm{BAO}}$ for an IBIS-like imaging survey, as a function of number density and survey area. These forecasts assume no interlopers, i.e., $f_{\rm int}=0$, though we consider their impact in \S \ref{sec:interlopers}.}
    \label{fig:forecast_ibis}
\end{figure}

In Figure \ref{fig:forecast_ibis}  we show Fisher forecast constraints for $\alpha_{\rm BAO}(z_{\rm eff}=2.8)$, combining information from all LAE auto-/cross-spectra. We start by assuming an idealized configuration with no interlopers. 
As the clustering measurements are shot noise-dominated, $\sigma(\alpha_{\rm BAO})$ scales roughly with $A_{\rm surv}^{-1/2}$ and $\bar{n}^{-1/2}$. Nonetheless, we find the position of the BAO feature can be recovered to better than 5\% assuming modest survey configurations ($A_{\rm surv}\geq 2000$ deg$^2$, $\bar{n}_{\rm LAE} \geq 625$ deg$^{-2}$). For configurations matching the desired 5000 deg$^2$ of IBIS coverage, $\sigma(\alpha_{\rm BAO}) < 3\%$, and assuming optimistic tracer density ($\overline{n}_{\rm LAE}=1250$ deg$^{-2}$) the forecast precision reaches 1.6\%.

We present constraints on $\alpha_{\rm BAO}$ that are marginalized over the spline nuisance parameters discussed in \S \ref{sec:splines} with fixed $\Delta \ell_{\rm spline}=350$. Compared to idealized Fisher forecasts without splines, our marginalized estimates of $\sigma(\alpha_{\rm BAO})$ are 2.5\% larger (as a percentage of the unmarginalized errors), reflecting minimal degradation from degeneracy between components. The fractional error degradation for per-redshift bin measurements is slightly higher, ranging from $4-6\%$ in the first three bins and $10-15\%$ for the last two.


\begin{figure}
    \centering
    \includegraphics[width=\linewidth]{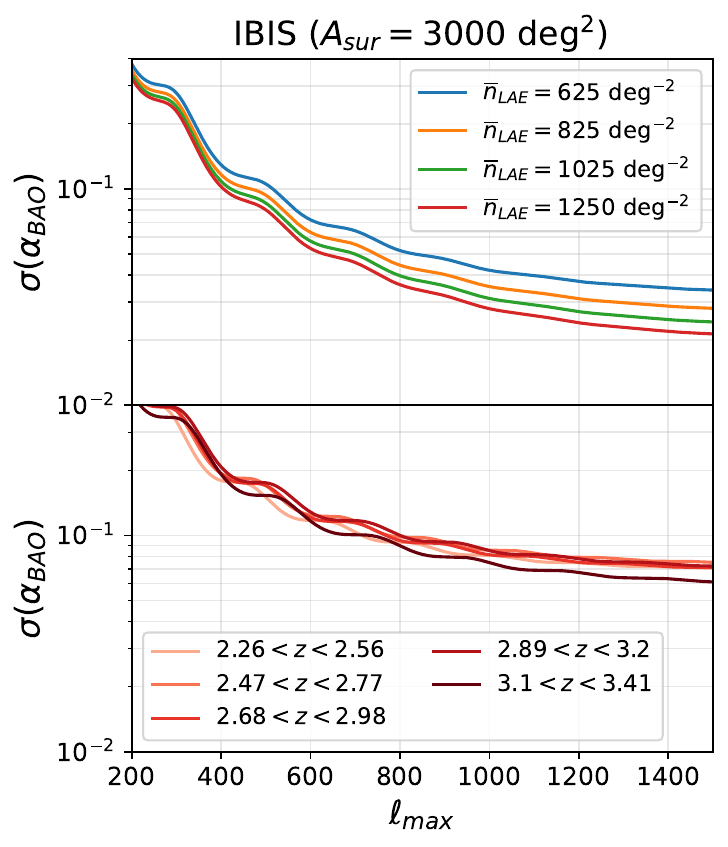}
    \caption{Dependence of $\sigma(\alpha_{\textrm{BAO}})$ on maximum angular multipole $\ell_{\rm max}$. To illustrate the scale dependence of the BAO information we show single-parameter forecasts for $\sigma(\alpha_{\rm BAO})$ (i.e., not marginalized over spline parameters). The top panel shows the combined constraints from all five redshift bins, for a range of tracer densities. The bottom panel shows the same but for individual redshift bins and a single tracer density ($\bar{n}_{\rm LAE}=825$ deg$^{-2}$). We assume no interlopers for these comparisons (see \S \ref{sec:interlopers}).}
    \label{fig:sigma_alpha_vs_ellmax}
\end{figure}

In Figure \ref{fig:sigma_alpha_vs_ellmax} we show the dependence of $\sigma(\alpha_{\rm BAO})$ on the maximum multipole $\ell_{\rm max}$ used in the Fisher forecasts. We find that including scales $\ell \gtrsim 1000$ is necessary to capture the majority of the BAO signal, though the information quickly diminishes beyond this point given the exponential damping of higher-$\ell$ BAO wiggles.

\subsection{Sensitivity to galaxy bias}

While our fiducial bias model is a smoothly varying function of the inferred biases from $z=2.5$ and $z=3.0$ \citep{ebina_ibis}, the measurements are currently uncertain and will depend on achieved photometric depth and sample selection criteria. To explore the sensitivity of our results to changes in the tracer bias, we consider three cases with redshift-independent biases $b=\lbrace 2.0, 2.25, 2.5 \rbrace$. For $A_{\rm surv}=5000$ deg$^{-2}$ and $\bar{n}_{\rm LAE}=825$ deg$^{-2}$, we find that $\sigma(\alpha_{\rm BAO})$ ranges from 3\% ($b=2.0$) to 2.2\% ($b=2.5$), reflecting $\pm 15\%$ relative variations in $\sigma(\alpha_{\rm BAO})$ about the intermediate bias case $b=2.25$. This suggests that variations in the bias at the level of current observational uncertainties have a modest impact on the forecast BAO precision. The actual LAE biases will depend on the details of the selected tracer sample and will be characterized with a subset of the primary survey.

\subsection{Interloper fraction}
\label{sec:interlopers}

Given challenges in the selecting faint, high-redshift LAEs, it is inevitable that some fraction of samples from imaging will contain interlopers. Interlopers that are uncorrelated with the target population dilute the measured signal, ultimately degrading BAO precision at fixed $\bar{n}_{\rm targ}$. To model the impact of interlopers in the imaging samples, we vary the interloper fraction $f_{\rm int}$ in our forecasts, retaining the initial source density while scaling $C_{\ell} \to C_{\ell}(1-f_{\rm int})^2$. In doing so, we assume that interloper sources have either low bias or are spread across a much wider $dN/dz$, such that the impact of interlopers is to increase the ratio between shot noise and recovered clustering signal.

\begin{figure}
    \centering 
    \includegraphics[width=\linewidth]{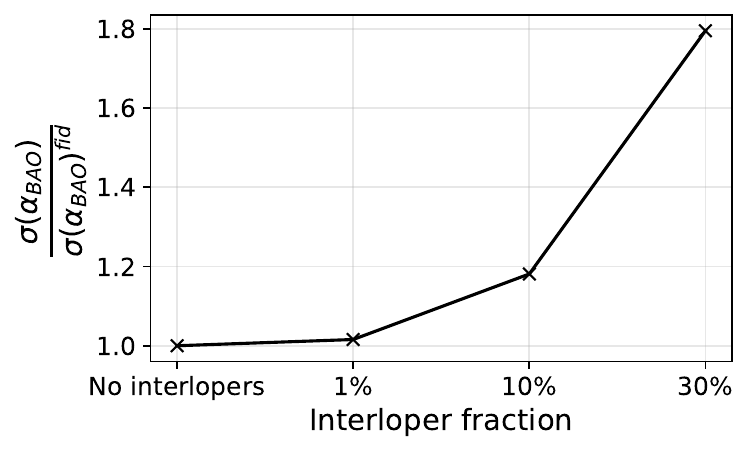}
    \caption{Degradation in $\sigma(\alpha_{\rm{BAO}})$ due to interlopers in the target sample, with varying interloper fraction.}
    \label{fig:ibis_interloper_degrade}
\end{figure}

We show the degradation of $\sigma(\alpha_{\rm BAO})$ relative to the case of no interlopers in Fig. \ref{fig:ibis_interloper_degrade}. For the range of IBIS LAE densities considered, the degradation due to higher interloper fractions does not vary significantly with assumed $\bar{n}_{\rm LAE}$. However, for denser surveys with $\bar{n}P\gg 1$ the tradeoffs become more nuanced, with less severe degradation in $\sigma(\alpha)$ as $\bar{n}$ increases.  


\section{Looking ahead to Stage-V surveys}
\label{sec:stageV}

We now explore more ambitious forecasts for high-redshift BAO measurements using the targeting from planned Stage-V surveys.
\citet{specS5} details a roadmap toward 3D clustering measurements over larger sky area that anticipates Rubin medium-band imaging (along with other alternatives) for target identification and improved instrumentation for spectroscopic follow-up. The fiducial specifications from \cite{specS5} include: seven MB filters spanning $3750 < \lambda < 5500$ \AA\ (corresponding to $2.1<z<3.5$), each with $\Delta \lambda \approx 250$ \AA, a total LAE target density of 3000 deg$^{-2}$, and survey area covering 11000 deg$^2$.

\begin{figure}
    \centering
    \includegraphics[width=\linewidth]{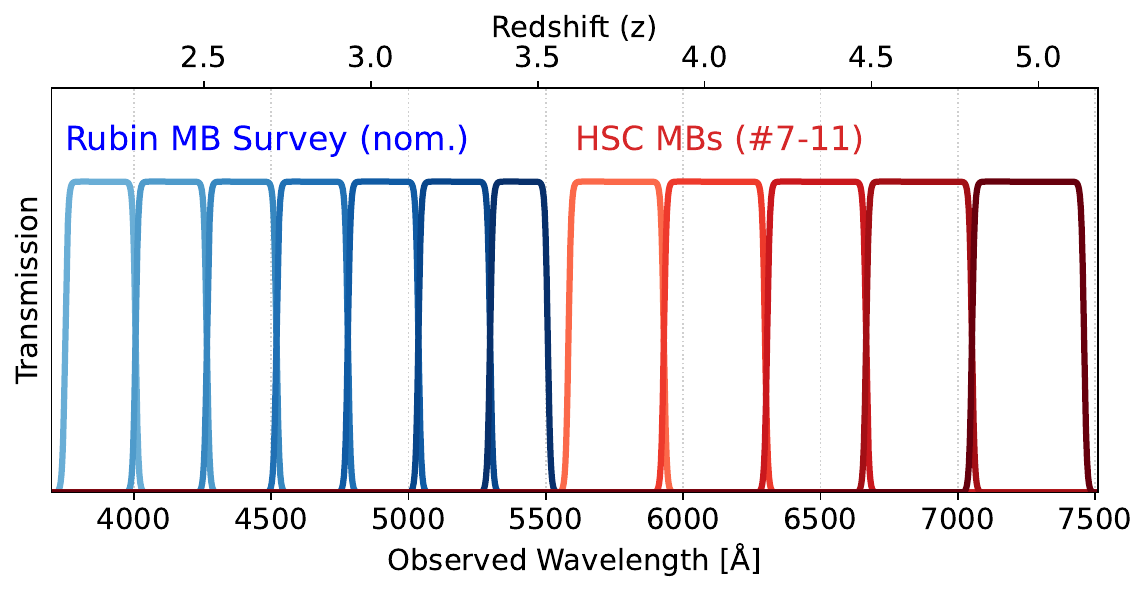}
    \caption{Assumed filter profiles for the planned Stage-V MB imaging (blue curves) and an extended, longer wavelength HSC MB-like configuration (red). The corresponding Lyman-$\alpha$ selection extends to $z=5.2$ (i.e., $\lambda=(1+z)1216$ \AA). We assume $dN/dz$ distributions that follow the normalized filter profiles.}
    \label{fig:s5_mediumband}
\end{figure}

We also consider a set of filters spanning $5500 <\lambda_{\rm obs}<7500$ \AA\ that track Ly-$\alpha$ over the extended range $3.5 < z < 5.2$. We assume $dN/dz$ distributions that follow the HSC medium-band survey filters\footnote{\url{https://subarutelescope.org/Instruments/HSC/sensitivity.html}}, which have $R\sim 20$ spectral resolution. In Figure \ref{fig:s5_mediumband} we show both the nominal and extended medium band filter profiles used in our forecasts. We restrict our forecasts to $z<5.2$, given larger model uncertainties on the clustering at $z>5$ and more severe observational challenges for ground-based measurements at $\lambda_{\rm obs}>7500$ \AA\ (e.g., absorption from water vapor and stronger OH airglow contributions).

There are a number of dedicated high-redshift LAE studies in the literature \citep{ouchi_2008, kovac_2007, khostovan_2019}, with some variation in the recovered large-scale LAE bias. We use the parametric formulae from \citep{ebina_white_highz}, which are derived from fits to a compilation of existing measurements, to model $b(z)$ up to $z=5$ and as a function of line flux,
\begin{equation}
    b(f_{\mathrm{lim}},z) = A(f_{\mathrm{lim}})(1+z)+B(f_{\mathrm{lim}})(1+z)^2,
\end{equation}
with
\begin{align*}
    A(f_{\mathrm{lim}}) &= 0.457 - 1.755(\log_{10}f_{\mathrm{lim}}+17) + 0.720(\log_{10}f_{\mathrm{lim}}+17)^2; \\
    B(f_{\mathrm{lim}}) &= 0.012 + 0.318(\log_{10}f_{\mathrm{lim}}+17)+0.043(\log_{10}f_{\mathrm{lim}}+17)^2.
\end{align*}
For both medium-band survey configurations, we assume a conservative interloper fraction of 30\%.

\subsection{Rubin-II medium-band survey}
\label{sec:rubinmb}

In order to achieve a constant comoving density, the target line fluxes for LAE detections (6$\sigma$) in Table 4 of \cite{specS5} vary from $\sim 1.2\times 10^{-16}$ erg cm$^{-2}$ s$^{-1}$ for the short wavelength bands down to $4\times 10^{-17}$ erg cm$^{-2}$ s$^{-1}$ for the redder bands. Given these line flux targets, the modeled biases for the seven medium band samples are $b=\lbrace 2.2, 2.4, 2.5, 2.7, 2.9, 3.0, 2.9\rbrace$ for $2.1 < z < 3.5$. 

\begin{figure}
    \centering
    \includegraphics[width=\linewidth]{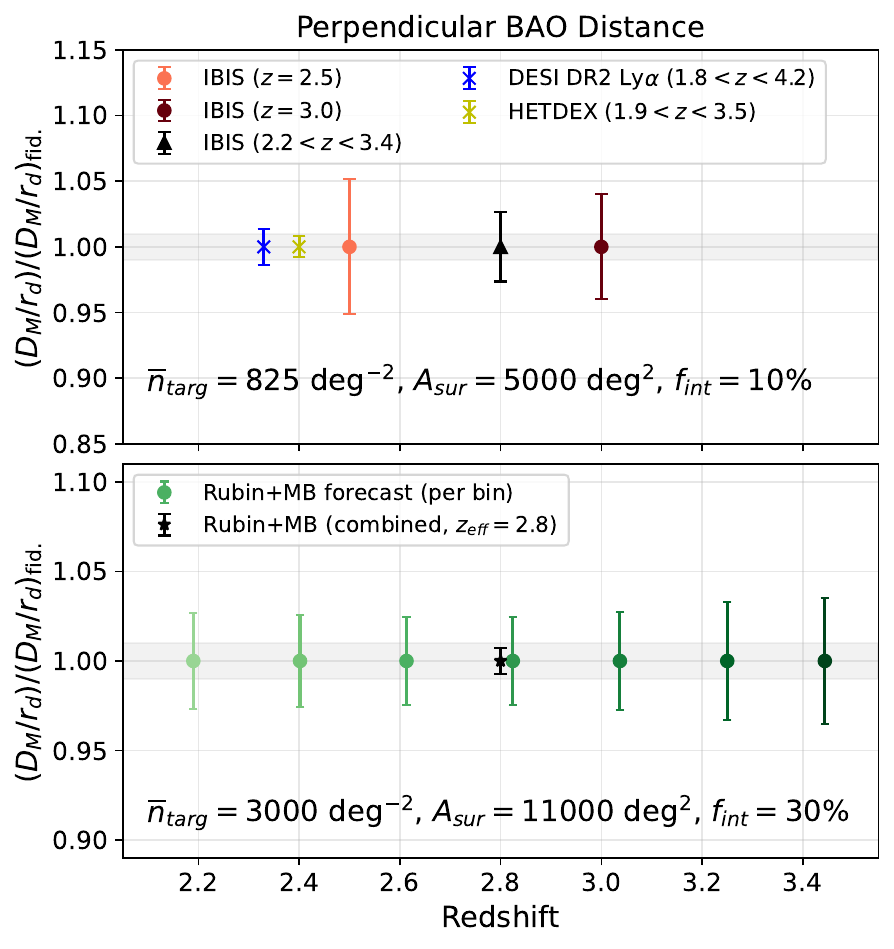}
    \caption{Forecast sensitivity to the perpendicular BAO distance, alongside DESI DR2 measurements with the Lyman-$\alpha$ forest \citep{desi_dr2_bao} and HETDEX forecasts \citep{hetdex_gebhardt}. The top panel shows a realistic configuration for IBIS ($\bar{n}_{\rm targ}=825$ deg$^{-2}$, $f_{\rm int}=0.1$, $A_{\rm surv}=5000$ deg$^2$), while the bottom panel shows forecasts for a Stage-V survey with \emph{Rubin}-like medium band (MB) imaging. The grey shaded bands indicate $\pm 1\%$ precision.}
    \label{fig:dm_ratio}
\end{figure}

We compare the results of these forecasts with our nominal IBIS constraints from \S \ref{sec:forecasts} in Fig. \ref{fig:dm_ratio}. While our IBIS forecasts predict 4-5\% precision for two redshift bins centered at $z=2.5$ and $z=3.0$, the per-bin \emph{Rubin} MB-like forecasts range from 2.5\% to 3.5\%, and reach 0.7\% precision when combining all seven LAE samples. We find that the highest redshift bins have a slightly larger covariance between BAO and the spline broadband model, explaining the mild degradation in uncertainties. 

\subsection{HSC extended medium band survey}

For our HSC-like MB configuration, we assume a constant line flux limit $f_{\mathrm{lim}}=4\times 10^{-17}$ erg cm$^{-2}$ s$^{-1}$ , which corresponds to LAEs with $b = \lbrace 3.3, 3.9, 4.5, 5.2, 6.0\rbrace$ across redshifts $3.5 < z < 5.2$. We also extend to $\ell_{\rm max}=2000$ in order to saturate the BAO information.

Pushing the targeting program to $3.5<z<5.2$ faces challenges including the detection of fainter tracers (at fixed luminosity) in imaging, the presence of time-variable airglow features (e.g., [OI], the sodium doublet (Na D), and OH), and higher Zodiacal light continuum foregrounds, some of which may also affect spectroscopic redshift efficiency if follow-up is pursued. Furthermore, published number densities for LAEs at $z>3.5$ disagree by a factor of a few, making it difficult to reliably forecast LAE abundances for fixed $f_{\rm lim}$. These considerations are beyond the scope of this work -- instead, we use our forecasting methodology in order to ask what tracer densities are required to deliver usable BAO constraints.  

\begin{figure}
    \centering
    \includegraphics[width=\linewidth]{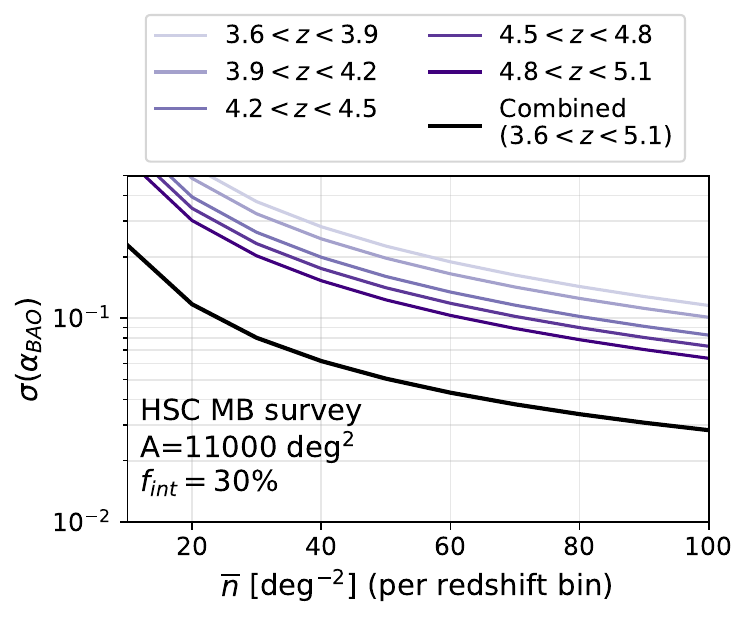}
    \includegraphics[width=\linewidth]{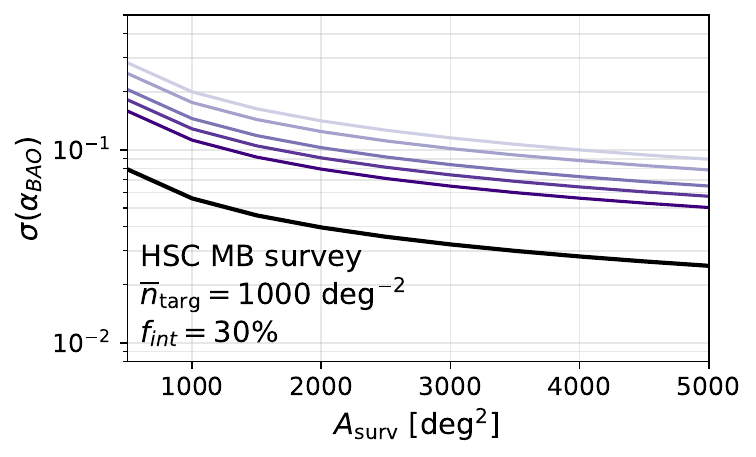}
    \caption{High redshift ($3.6<z<5.1$), HSC MB-like forecasts as a function of tracer density (top) and survey area (bottom). In each panel the colored curves indicate the per redshift bin forecasts while the black curve indicates the combined sensitivity to the BAO scale with $z_{\rm eff}=4.35$.}
    \label{fig:hscmb_vs_nbar}
\end{figure}

In Figure \ref{fig:hscmb_vs_nbar} we show the forecast constraints on $\alpha_{\rm BAO}$ for individual and combined redshift bins defined by the HSC-MB filters. This is explained by the assumed increasing LAE bias with redshift. With a footprint matching that of the previous Rubin MB specifications (see \S \ref{sec:rubinmb}), we find that the required number density to surpass $\sigma(\alpha_{\rm BAO}) < 10\%$ varies from $\bar{n} \approx 100$ deg$^{-2}$ for the lowest redshift bin ($3.6<z<3.9$). This decreases to $\overline{n}_{\rm min}\sim 60$ deg$^{-2}$ for the $4.8<z<5.1$ bin, which is explained by increasing $b(z)$. Combining the five MB measurements, $\sigma(\alpha_{\rm BAO})$ ranges from 11\% ($\overline{n}_{\rm tot}=100$ deg$^{-2}$) down to 3\% ($\overline{n}_{\rm tot}=500$ deg$^{-2}$). 

As the measurements are shot noise-dominated, one can trade survey area and tracer density for a given target BAO precision. Furthermore, to obtain LAE samples at $z>3.5$ with reasonable biases for cosmological analysis, one needs to achieve flux limits $f_{\rm lim}\sim 4\times 10^{-17}$ erg cm$^{-2}$ s$^{-1}$, at which point the number densities implied from existing measurements are $\overline{n}\gtrsim 2\times 10^{-4}$ h$^3$ Mpc$^{-3}$, i.e., higher than we just considered. Taking these considerations into account, we also show forecasts for configurations with constant comoving density $\overline{n}=2\times 10^{-4}$ h$^3$ Mpc$^{-3}$ (corresponding to $\overline{n}_{\rm tot}\simeq 1000$ deg$^{-2}$ projected over $3.6<z<5.1$) and vary the survey area across $A_{\rm surv}\in [500, 5000]$ deg$^2$. We find that configurations with $A_{\rm surv}>2000$ deg$^2$ can reach a combined precision on $\alpha_{\rm BAO}$ surpassing 3\%, while BAO measurements for individual redshift bins become generally feasible for $A_{\rm surv}>4000$ deg$^2$.

%% file: conclusion.tex
\section{Conclusion}
\label{sec:conclusion}
We have presented forecasts of transverse baryon acoustic oscillation (BAO) constraints from high-redshift galaxy samples selected with medium-band imaging, motivated by the forthcoming DESI-II program and associated medium band imaging through IBIS to target Lyman-$\alpha$ emitters spanning $2.2 < z < 3.5$. Using a range of possible specifications for the IBIS LAE survey informed by \cite{ebina_ibis}, we find that the angular BAO distance scale, combining all MB samples, can be measured to better than $5\%$ for $A_{\rm surv}>2000$ deg$^2$ and with $2-3\%$ precision for coverage over $5000$ deg$^2$ and tracer densities $\bar{n}>2\times 10^{-4}$ h$^3$ Mpc$^{-3}$. 

We then consider a more ambitious, sample variance-limited LAE survey that utilizes \emph{Rubin}-like MB imaging anticipated for the Stage-V spectroscopic survey, and predict a combined precision of $0.7\%$ on the angular BAO scale at $z_{\rm eff}=2.8$ when combining seven redshift bins across $2.1<z<3.5$. Furthermore, we find that measurement of the BAO feature at even higher redshifts ($3.5<z<5.2$) is possible using longer wavelength HSC-like MBs for LAE selection. A full characterization of observational challenges and model-based uncertainties at these redshifts is beyond the scope of this work, however with deeper pilot surveys it should be possible to refine the existing forecasts and reassess the plausibility of a dedicated $z>3.5$ LAE survey.

The results presented here support the general feasibility of extracting BAO distance information from medium-band imaging surveys with partial or incomplete spectroscopic follow-up. With the main IBIS survey underway, larger samples and refined selections will determine the final achieved number densities and biases of targeted LAEs. Future work may involve more detailed treatments of observational systematics needed for the angular clustering measurements. We likewise did not explore the potential complementarity of our LAE constraints with other overlapping tracers of structure such as the Lyman-$\alpha$ forest using quasars, which have fine radial resolution but do not trace large angular scales as well. 

\section*{Acknowledgements}

The authors thank Haruki Ebina and David Schlegel for guidance regarding the target properties of the IBIS medium band samples and for comments on the manuscript. R.M.F. is partially supported by a Berkeley Center for Cosmological Physics (BCCP) fellowship. M.W. was funded by the DOE.